\newcommand{\qB}{\ensuremath{\theta_B}}
\newcommand{\normvar}[1]{\ensuremath{\left(\Delta \left(#1\right)\right)^2}}
\newcommand{\meanJ}{\ensuremath{|\langle\vc{J}\rangle|}}
\newcommand{\vc}[1]{\ensuremath{\mathbf{#1}}}
\newcommand{\bra}[1]{\ensuremath{\left\langle {#1} \right|}}
\newcommand{\ket}[1]{\ensuremath{\left|  #1 \right\rangle}}

\documentclass[aps,prl,showpacs,amsmath,amssymb, twocolumn]{revtex4}
\pdfoutput=1
\usepackage{graphicx}
\usepackage{hyperref}
\usepackage{bibunits}

\begin{document}
\begin{bibunit}
\title{Conditional Spin-Squeezing of a Large Ensemble via the Vacuum Rabi Splitting}

\author{Zilong Chen}
\author{Justin G. Bohnet}
\author{Shannon R. Sankar}
\author{Jiayan Dai}
\author{James K. Thompson}
\affiliation{JILA, NIST, and Department of Physics, University of Colorado, Boulder, Colorado 80309-0440, USA}
\date{\today}

\begin{abstract}
We use the vacuum Rabi splitting to perform quantum nondemolition (QND) measurements that prepare a conditionally spin-squeezed state of a collective atomic psuedo-spin.   We infer a 3.4(6)~dB improvement in quantum phase estimation relative to the standard quantum limit for a coherent spin state composed of uncorrelated atoms.    The measured collective spin is composed of the two-level clock states of nearly $10{^6}$ $^{87}$Rb atoms confined inside a low finesse $F = 710$ optical cavity.  This technique may improve atomic sensor precision and/or bandwidth, and may lead to more precise tests of fundamental physics.
\end{abstract}


\maketitle

Large ensembles of uncorrelated atoms are extensively used as precise sensors of time, rotation, and gravity, and for tests of fundamental physics \cite{LZC08,GBK97, MOT08,WHS02}.  The quantum nature of the sensors imposes a limit on their ultimate precision.  Larger ensembles of $N$ atoms can be used to average the quantum noise as $1/\sqrt{N}$, a scaling known as the standard quantum limit.  However, the ensemble size is limited by both technical constraints and atom-atom collisions--a fundamental distinction from photon-based sensors.    Learning to prepare entangled states of large ensembles with noise properties below the standard quantum limit will be key to extending both the precision \cite{ASL04} and/or bandwidth \cite{ABK04} of atomic sensors.  More broadly, the generation and application of entanglement to solve problems is a core goal of quantum information science being pursued in both atomic and solid state systems.

In this Letter, we utilize the tools of cavity-QED to prepare an entangled ensemble with a $3.4(6)$ dB improvement in spectroscopic sensitivity over the standard quantum limit.  The method does not require single particle addressability and is applied to a spectroscopically large ensemble of  $N= 7\times 10^5$ atoms using a single $<200~\mu$s operation.  The gain in sensitivity is spectroscopically equivalent to the enhancement obtained had we created $>10^5$ pairs of maximally entangled qubits, demonstrating the power of a top-down approach for entangling large ensembles. The probing of atomic populations via the vacuum Rabi splitting is also of broad interest for non-destructively reading out a wide variety of both atomic and solid state qubits. 
 
The large ensemble size is a crucial component.  Entangled states of cold, neutral atoms are unlikely to impact the future of quantum sensors and tests of fundamental physics unless the techniques for generating the states are demonstrated to work for the  $10^4$ to $10^7$ neutral atom ensembles typically used in primary frequency standards \cite{HJD05} and atom interferometers \cite{GBK97,WHS02}.   

The approach described here allows quantum-noise limited readout of a sensor with $<0.2$ photon recoils/atom, producing little heating of the atomic ensemble.  Applied to a state-of-the-art optical lattice clock, the resulting enhanced measurement rates will suppress  the dominant aliasing of the local oscillator noise \cite{LZC08,LWL09}.   

The gain in spectroscopic sensitivity demonstrated here  is far from the fundamental Heisenberg limit which scales as $1/N$, a limit approached by creating nearly maximally entangled states of  2 to 14 ions \cite{LBS04,MRK01,MSB10}.  However, the gain here relative to the standard quantum limit is comparable to these experiments.    Ensembles of $N\approx 10^3$ atoms have been spin-squeezed by exploiting atom-atom collisions within a Bose-Einstein Condensate \cite{GZN10,RBL10}, however these systems face the significant challenge of managing systematic errors introduced by the required strong atomic interactions.  

Spin-squeezed states can also be prepared with atom-light interactions that generate effective long range interactions on demand. In the approach followed here \cite{KBM98}, light is used to perform a measurement that projects the ensemble into a conditionally spin-squeezed state, as shown for clock transitions with laser cooled atoms in free space (3.4~dB at $N=1.2\times10^5$ \cite{AWO09}) and in a cavity (3.0~dB at $N = 3.3\times10^4$ \cite{SLV10}).  Conditional two-mode squeezing of a room temperature vapor of $N=10^{12}$ atoms enabled magnetometry with 1.5~dB of spectroscopic enhancement and an increased measurement bandwidth \cite{WJK10}.  A  non-linear atom-cavity system also generated $5.6$ dB of spin-squeezing  at $N = 3\times 10^4$ atoms \cite{LSV10}.

The work we present here is unique in that we probe the atomic ensemble in the on-resonance regime of strong collective coupling cavity-QED.   By doing so, we hope to counter a commonly held view that the quality of a coherence-preserving quantum nondemolition (QND) measurement is fundamentally linked to the probe's large detuning from atomic resonance.   Instead, it is the magnitude of the collective cooperativity parameter $N C$ (equivalent to the optical depth for a free space experiment) that sets the fundamental quality of the QND measurement \cite{ABK04}.   In the context of free space measurements, detuning from resonance creates little enhancement in sensitivity once the detuned optical depth falls below one.  Using an optical cavity to enhance the cooperativity parameter has the potential to allow similar results to free space experiments, but at atomic densities lowered by a factor of order the cavity finesse.   

\begin{figure}[t]
\includegraphics[width=3.4in]{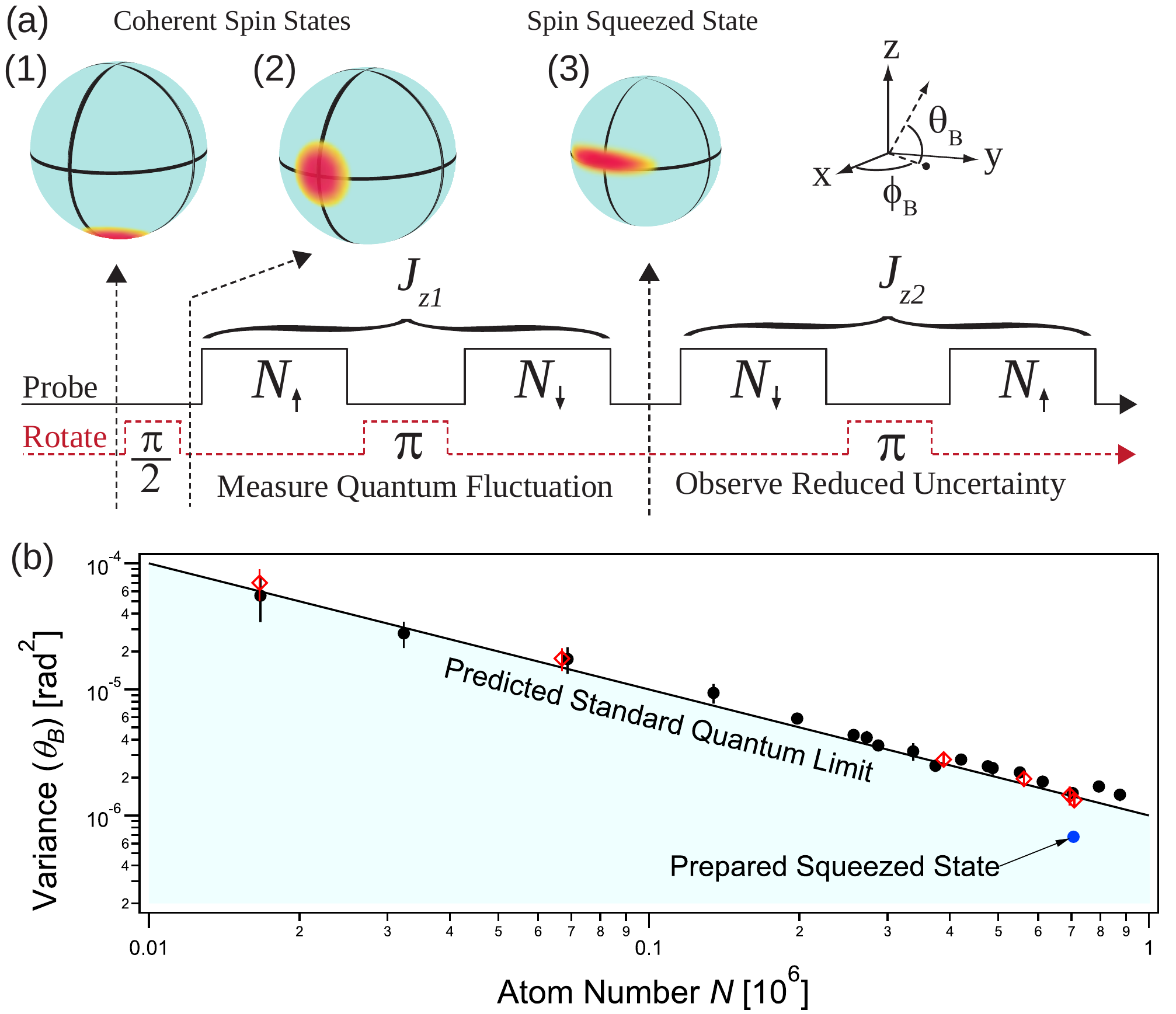}
\caption{(color online).  (a) The quantum noise of a collection of psuedo-spin $1/2$ atoms can be represented as a classical probability distribution for the collective spin or Bloch vector with length $J_{max} = N/2$. Probing the number of atoms in spin up and spin down measures the projection $J_z = (N_\uparrow - N_\downarrow)/2$, as well as $J_{max}$. For states near the equator, the polar angle is $\qB \approx J_z/J_{max}$.  The population measurement sequence consists of composite microwave rotations (dashed line) and QND population measurements (solid line).  The evolution of the ensemble is represented by Bloch spheres 1, 2, and 3. The QND measurement projects the ensemble into a conditionally-squeezed state (3), which is verified with the second measurement.   (b) The observed variance of $\qB$ versus atom number confirms the predicted standard quantum limit.  Two different rotations (circles and diamonds) are used to constrain added noise from the rotations \cite{OSM}.}
\label{Fig1}
\end{figure}

Each atom in the  ensemble can be represented as a pseudo spin-1/2, with the quantity to be measured (i.e. energy splitting, acceleration, etc.) represented by the magnitude of an effective magnetic field that causes the total spin or Bloch vector to precess \cite{WBI94}.  Quantum mechanics sets a fundamental limit on our ability to measure the precession angle $\phi$ and hence infer the value of the effective magnetic field.  For an ensemble of uncorrelated spins, the quantum phase uncertainty is $\Delta \phi = 1/\sqrt{N}$, and is referred to as the standard quantum limit  for a coherent spin-state (CSS). This uncertainty can be more generally visualized  as a classical probability distribution of possible positions of the tip of a classical vector on the surface of a Bloch sphere, as shown in Fig. \ref{Fig1}.  This noise is equivalent to projection noise arising from measurement-induced collapse into spin up or down \cite{WBI94}.

A quantum particle's position can be determined with unlimited precision at the expense of knowing its momentum.  We demonstrate QND measurements that reduce the uncertainty in the polar angle $\theta_B$ describing the Bloch vector, at the expense of quantum back-action  appearing in the azimuthal angle $\phi_B$. 

\begin{figure}
\includegraphics[width=3.4in]{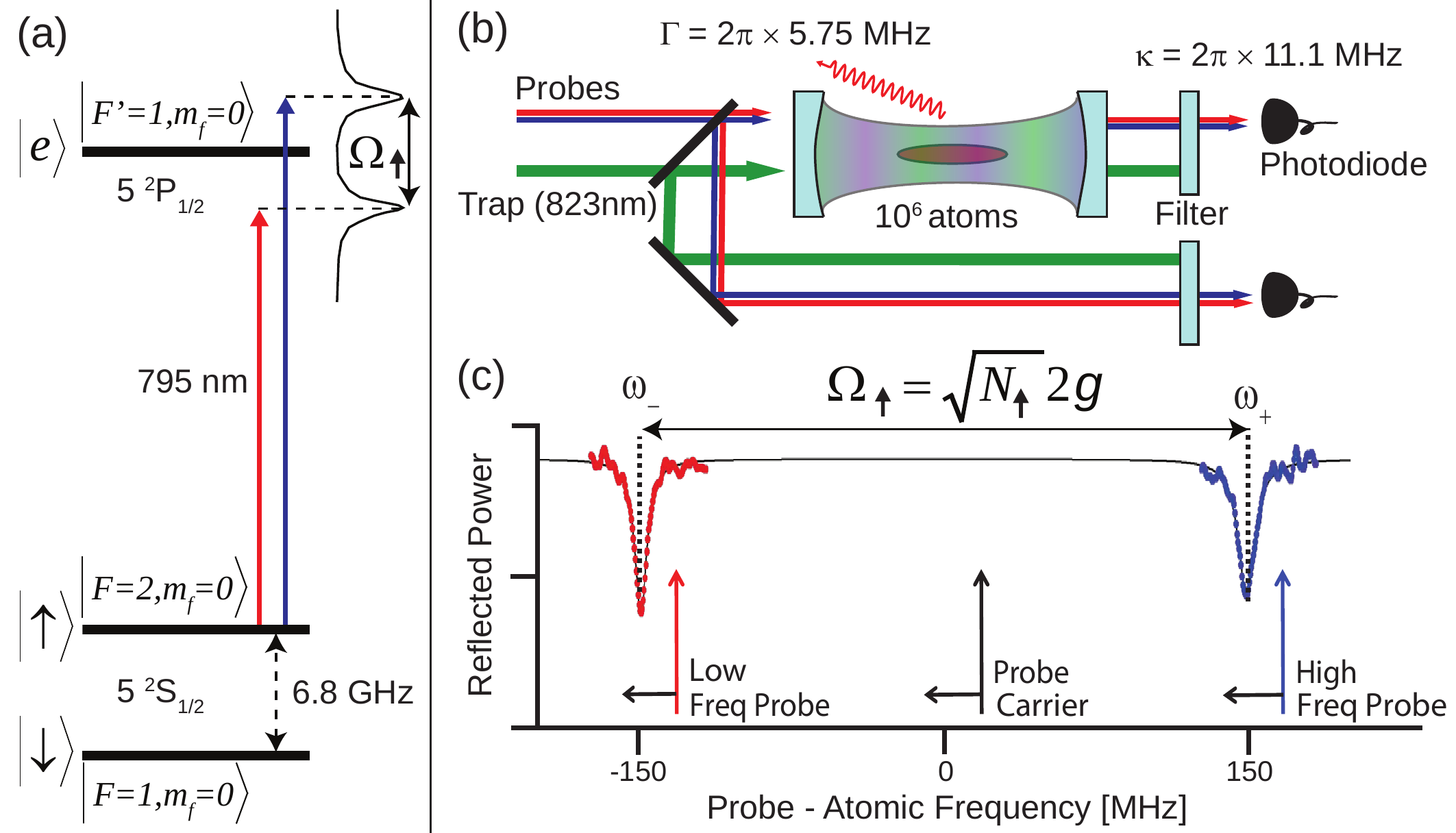}
\caption{(color online). (a)  The clock states \ket{\uparrow} and \ket{\downarrow} form a pseudo spin-$\frac{1}{2}$ system. The population $N_\uparrow$ is measured by probing near the \ket{\uparrow} to \ket{e} transition, which couples to a degenerate cavity mode, creating a collective vacuum Rabi splitting $\Omega_{\uparrow}$. (b) An ensemble of $10^6$ atoms are tightly confined within the $\text{TEM}_{00}$ mode using a 1D intra-cavity optical lattice at wavelength 823~nm.  The cavity and atomic decay rates are $\kappa$ and $\Gamma$ respectively. A heterodyne interferometer (not shown) is used to probe the atom-cavity resonances.  The collective nature of the Rabi splitting prevents loss of coherence as atoms remain in a superposition after the measurement. (c)  The vacuum Rabi splitting $\Omega_{\uparrow}= \omega_+ - \omega_- $   is measured by simultaneously scanning the probe frequencies across the upper and lower atom-cavity resonances $\omega_\pm$.   The size of the fitted splitting determines the population  in $\ket{\uparrow }$ from $N_\uparrow =( \Omega_\uparrow/2 g)^2$.  A single scan requires $\approx70~\mu$s.}
\label{Fig2}
\end{figure}

A pseudo spin-$\frac{1}{2}$ system is formed by the clock states $\ket{\uparrow} \equiv \ket{F=2, m_F=0}$ and $\ket{\downarrow} \equiv \ket{F = 1, m_F =0}$ of  $^{87}$Rb (Fig.~\ref{Fig2}a).  The ensemble of $N$ particles is described by a collective Bloch vector $\vc{J} = \sum_i \vc{j}_i$ ($\vc{j}_i$ is the spin of the $i$-th particle). In the fully symmetric manifold, the Bloch vector has length $J = J_{max} = N/2$ and $\langle \vc{J}^2 \rangle = J(J+1)$. The z-component of the Bloch vector is proportional to the population difference between the $\ket{\uparrow}$ and $\ket{\downarrow}$ states, $J_z = (N_{\uparrow} - N_{\downarrow})/2$.  In our experiments, the Bloch vector is prepared through a combination of optical pumping and microwave induced rotations in the state   $\left<\vc{J}\right> =  \hat{x} N /2$.  The polar angle is determined by the measured populations $\theta_B \approx J_z/\left |\left<\vc{J}\right>\right | =  (N_{\uparrow} - N_{\downarrow})/(2\left |\left<\vc{J}\right>\right |)$.  We show that reduced uncertainty states with respect to $\theta_B$ can be prepared by first demonstrating that  $J_z$ can be measured with precision better than the CSS noise $\Delta J_{zCSS} = \sqrt{N}/2$, and then by demonstrating that the length of the Bloch vector $\left |\left<\vc{J}\right>\right |$ is only slightly reduced.

The atoms are trapped inside an optical cavity tuned to resonance with the \ket{\uparrow} to \ket{e} $\equiv$ \ket{5^2P_{1/2}, F=1, m_F=0} optical transition with wavelength $\lambda= 795$~nm  (see Fig. \ref{Fig2}b).  To account for lattice sites where atoms only weakly couple to the probe mode, we follow the procedure of Ref.~\cite{SLV10} by defining effective parameters, $N$ and $g$, such that $\Delta J_{zCSS}$ remains $\sqrt{N}/2$ \cite{OSM}. The effective single particle vacuum Rabi frequency is  $2 g= 2 \pi \times 506(8)$~kHz \cite{OSM}. Coupling to the cavity mode is enhanced by using up to $N=7 \times 10^5$ so that the collective cooperativity parameter $N_{\uparrow} C \approx 1400$ is large.

\begin{figure}
\includegraphics[width=3.4in]{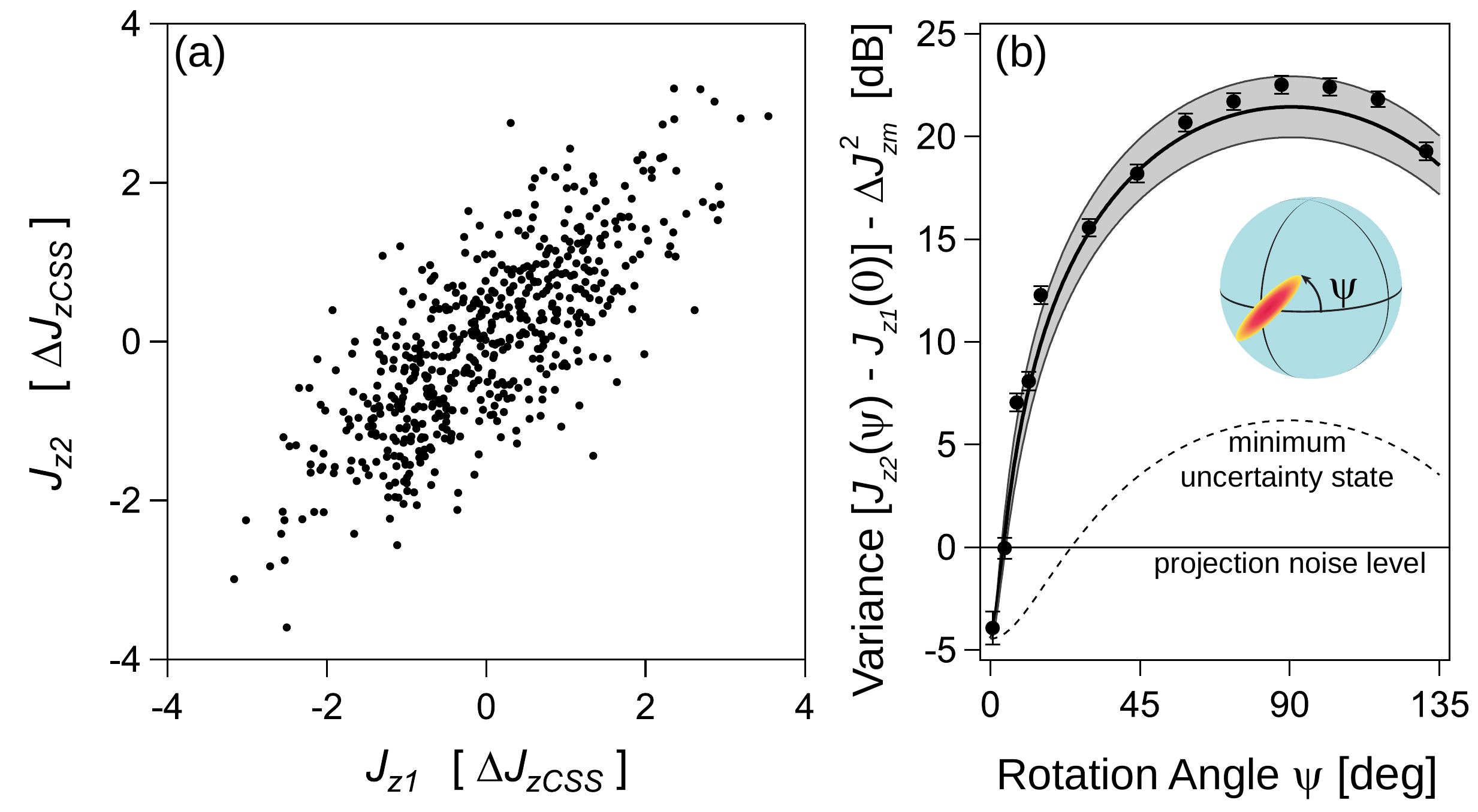}
\caption{(color online). (a) The first and second QND measurements of $J_z$ exhibit correlated fluctuations that arise largely due to projection noise.  (b) The noise in the back-action quadrature is observed by inserting a rotation through angle $\psi$ between the QND measurements $J_{z1}$ and $J_{z2}$.  The dashed curve is the calculated response for a minimum uncertainty squeezed state, while the solid curve with the gray 68\% confidence interval is the predicted back-action from the intracavity probe vacuum noise.}
\label{Fig3}
\end{figure}

The atomic population is determined from $N_{\uparrow} = \left(\Omega_{\uparrow}/2g\right)^2$, where  $\Omega_{\uparrow}$ is the collective vacuum Rabi splitting \cite{ZGM90}.  The bare cavity mode is dressed by the presence of atoms in state \ket{\uparrow} to generate two new resonances at frequencies  $\omega_\pm$ relative to the original cavity resonance (see Fig.~\ref{Fig2}c).  The measured splitting $\Omega_{\uparrow} = \omega_+ -\omega_-$ is only quadratically sensitive to detuning between the atomic and bare cavity resonances, relaxing technical requirements on cavity stability.  Requirements on laser frequency stability are also relaxed by simultaneously scanning the resonances using two probe frequency components generated by phase modulating a single laser.  

The other population $N_\downarrow$ is determined by first applying a microwave $\pi$-pulse that phase coherently swaps the populations between \ket{\downarrow} and \ket{\uparrow} (see Fig.~\ref{Fig1}a). The population of \ket{\uparrow} is then determined from the vacuum Rabi splitting with the results labeled $N_\downarrow$ and $\Omega_{\downarrow}$.  For scale, the predicted projection noise $\Delta J_{zCSS}$ would produce rms fluctuations in the vacuum Rabi splittings of $\Delta ( \Omega_{\uparrow} - \Omega_{\downarrow}) = \sqrt{2} g = 2\pi \times 358(6)$~kHz.

The predicted projection noise variance $(\Delta J_{zCSS})^2$ is confirmed to 2(6)\% by measurements of the variance of $J_{z1}$ versus atom number (see Fig.~\ref{Fig1}b).  Projection noise results in a linear dependence of the variance with atom number, whose magnitude is determined using low order polynomial fits.  The fitted linear contribution is $1.02 \pm 0.05 (\mathrm{stat}) \pm 0.04 (\mathrm{syst})$ times $(\Delta J_{zCSS})^2$.

We now demonstrate that repeated measurements of $J_z$ are correlated below the projection noise level $\Delta J_{zCSS}$.  A first measurement $J_{z1}$ estimates $J_z$ to a precision set primarily by the measurement noise $\Delta J_{zm}$, preparing a sub-projection noise state  when  $\Delta J_{zm} <\Delta J_{zCSS}$.  As shown in Fig. \ref{Fig3}, quantum projection noise plus added classical and detection noise causes fluctuations in the measured $J_{z1}$ from one trial to the next, but the fluctuations are partially correlated with a second measurement $J_{z2}$, allowing the quantum noise to be partially canceled in the difference $J_{z2} -J_{z1}$.

At $N_0 = 7.0(3) \times 10^5$ atoms and a  probe photon number of $M_0 = 1.9(1) \times 10^5$ per measurement of $J_z$, the variance of the difference of two QND measurements was $\normvar{J_{z2}- J_{z1} }/(\Delta J_{zCSS})^2 = -2.6(3)~$dB,  where a Bayesian estimator for $J_{z1}$ was applied.  Subtracting the measurement noise $\Delta J_{zm}$ of the second measurement  in quadrature gives a conservative estimate of the uncertainty in $J_z$  after the first QND measurement of $(\Delta J_{z})^2 /(\Delta J_{zCSS})^2  = -4.9(6)~$dB.  The noise $\Delta J_{zm}$ is determined from fluctuations in the difference of two time adjacent $N_{\downarrow}$ measurements \cite{OSM}.  Accounting for fluctuations in Raman scattering to other magnetic sub-levels does not change this result.

\begin{figure}
\includegraphics[width=3.4in]{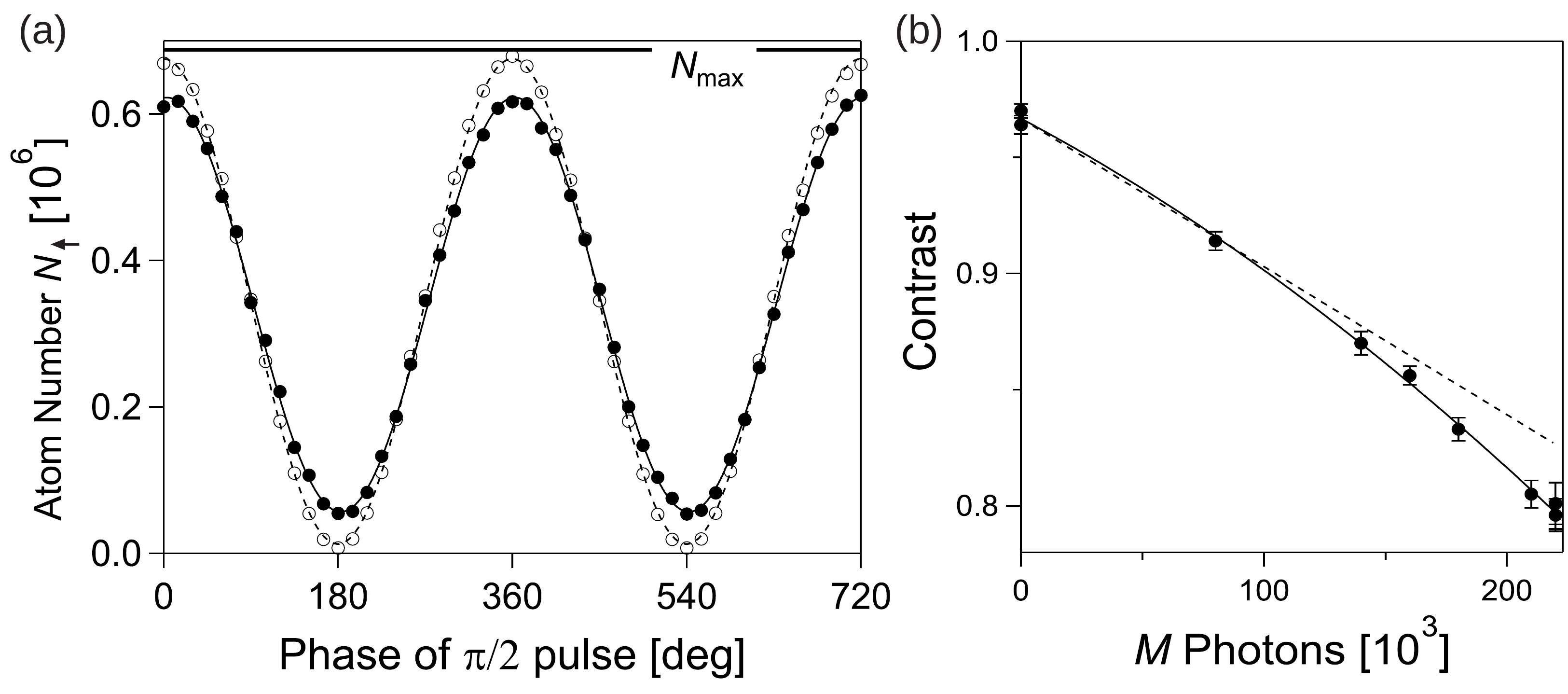}
\caption{(a) The degree of coherence remaining after the  measurement $J_{z1}$ is determined using the  sequence: ($\frac{\pi}{2}$)--(measure $N_{\uparrow}$)--($\pi$)--(measure $N_{\downarrow})$--($\frac{\pi}{2}$)--(measure $N_{\uparrow}$).  The sequence is repeated and the final measured value of $N_{\uparrow}$ is plotted versus the phase of the final $\frac{\pi}{2}$-pulse.  With no measurements (empty circles), the background contrast is $\mathcal{C}_i = 0.97(1)$.  Probing with $1.8\times10^5$ photons (filled circles), causes a small reduction in contrast despite a measurement sensitivity below the projection noise level.  (b) Measured contrast versus probe photon number (solid circles), second order polynomial fit (solid line), and the predicted contrast loss due to free space scattering alone  (dashed line).}
\label{Fig4}
\end{figure}

The unmeasured azimuthal angle $\phi_B$ is driven by fluctuating AC stark shifts arising from the intracavity probe vacuum noise (see Fig.~\ref{Fig3}b and Ref.~\cite{TVL08}.)  The measured and predicted quantum back-action noise levels are  22.3(1) dB and 21.4(1.5)~dB relative to projection noise respectively.  The back-action is larger than that of a minimum uncertainty squeezed state due to finite quantum and technical efficiencies in the probe detection process.

Reduction of spin noise alone does not allow improved quantum phase estimation unless the length of the Bloch vector $\meanJ$ is sufficiently unchanged.  The normalized length of the Bloch vector is measured by varying the polar angle $\theta_B$ from $ 0$ to $2 \pi$ and determining the contrast $\mathcal{C}= \meanJ/J_{max}$  from the resulting variation of the population $N_{\uparrow}$ (see Fig.~\ref{Fig4}).  Before the QND measurements, the contrast is $\mathcal{C}_{i} = 0.97(1)$, and the first QND measurement reduces the contrast to $\mathcal{C}_f = 0.82(2)$.  

Free space scattering of probe photons leads to collapse of the spin wavefunction of individual spins and is the dominant source of decoherence loss.  A rate equation analysis predicts the number of probe photons scattered into free space $M_{sc}$ is  related to the total number of probe photons $M$ by $M_{sc}/M = 0.41(1)$.  This prediction is in excellent agreement with the measured value  0.41(2) deduced by measuring the decrease in the vacuum Rabi splitting due to Raman scattering versus probe photon number.  If each scattered photon leads to the collapse of a single spin, then the fractional reduction in contrast is $6.4(3)\times10^{-7}\times M$ at $N_0$ atoms.  As shown in Fig.~\ref{Fig4}, the measured contrast versus probe photon number is well described by $\mathcal{C}_f = \mathcal{C}_i -  k_1M - k_2 M^2$. The fitted value, $k_1 = 5.5(7)\times10^{-7}$ per photon, is in good agreement with the prediction and confirms the fundamental role of free space scattering as the dominant source of decoherence.

The quadratic variation of $\mathcal{C}_f $, $k_2 = 1.0(3)\times10^{-12}$ per (photon)$^2$, arises from uncanceled inhomogeneous probe-induced light shifts that result in dephasing of the ensemble.   These light shifts are largely spin-echoed away with the $\pi$--pulse used to measure $J_{z1}$. The uncanceled dephasing arises from radial motion in the trap.  At fixed measurement precision, the magnitude of the dephasing increases linearly with probe detuning, making it easier to reach a scattering-dominated regime in this work compared to work in a far-detuned dispersive regime \cite{SLV10}. 

The ability to estimate the polar angle $\theta_B\approx J_z/\meanJ$ is largely set by the noise in $J_z$ and the signal size $\meanJ$. From Ref.~\cite{WBI94}, the directly observed spectroscopic gain is given by $\zeta_m^{-1}= \mathcal{C}_f^2 (\Delta J_{zCSS})^2  /(\mathcal{C}_i \normvar{J_{z2} - J_{z1}}) = 1.1(4)$~dB  below the standard quantum limit.  We infer the ability to prepare states with enhanced spectroscopic sensitivities of $\zeta_m^{-1}=  \mathcal{C}_f^2 (\Delta J_{zCSS})^2 /(\mathcal{C}_i  (\Delta{J}_{z})^2 ) = 3.4(6)$~dB.  

The coherence-preserving nature of the QND measurements here should also be contrasted with weak, sampled measurements in order to compare this work to fluorescence detection of an optically thin ensemble. The angle  $\qB$  could be estimated by extracting 15\% of the initially un-decohered atoms, and performing perfect state detection on this sub-ensemble.  The loss of signal would be the same as observed in our experiment,  but the sub-ensemble's estimate of $\qB$ would be $G= 13(1)$~dB noisier than the precision demonstrated using a collective measurement approach.  This reduction in noise can be described as arising from a noiseless amplifier of gain $G$ placed before a 15/85 atom beam splitter \cite{GLP98}.

In the future, this method can be extended to achieve greater violations of the standard quantum limit since many experimental aspects, such as cavity finesse and length, can be easily improved. Running wave cavities or commensurate optical lattices can be employed to create squeezed states appropriate for launching ensembles into free space for matter wave interferometry.

We thank A.~Hati, D.~Howe, K.~Lehnert, and L.~Sinclair for help with microwaves, and H.~S.~Ku, S.~Moses, and D.~Barker for early contributions. This work was supported by the NSF AMO PFC and  NIST.  Z.C. acknowledges support from A*STAR Singapore.

\end{bibunit}

\begin{bibunit}
\title{Conditional Spin-Squeezing of a Large Ensemble via the Vacuum Rabi Splitting:\\ Supporting Online Material}

\author{Zilong Chen}
\author{Justin G. Bohnet}
\author{Shannon R. Sankar}
\author{Jiayan Dai}
\author{James K. Thompson}
\affiliation{JILA, NIST, and Department of Physics, University of Colorado, Boulder, Colorado 80309-0440, USA }
\date{\today}


\maketitle

\subsection{Effective Coupling and Atom Number}

The atomic ensemble of $N_{tot}$ atoms confined in the 823~nm 1D optical lattice experiences inhomogeneous coupling to the incommensurate standing-wave probe mode at 795~nm, with some atoms essentially uncoupled to the probe mode.  Thus the standard quantum limit for the sub-ensemble of probed atoms is larger than that of the total ensemble $N_{tot}$.  To account for this inhomogeneous coupling, we follow the procedure of Refs.~\cite{SLV10,AWO09, LSV10}, and define effective atom numbers $N_{\uparrow}$, $N_{\downarrow}$ and an effective coupling $g$.  The definition of $N_{\uparrow,\downarrow}$ is chosen such that the variance of the quantum projection noise for a state $\langle \vec{J} \rangle = \hat{x} N/2$ is $(\Delta J_z)^2 = N/4$ as one would expect for an ensemble of $N=N_\uparrow + N_\downarrow$ uniformly coupled atoms.  The effective coupling is then chosen to produce the observed vacuum Rabi splittings for the probed sub-ensemble  $\Omega_{\uparrow, \downarrow} = \sqrt{N_{\uparrow, \downarrow}}(2 g)$.   These two requirements are equivalent to the following conditions that determine the effective parameters $N_\uparrow$ and $g$:  $\Big \langle \sum_{i=1}^{N_{\mathrm{tot}   }} \hat{P}_{\uparrow, i}(2g(\vec{r}_i))^2\Big \rangle  = N_\uparrow (2 g)^2$  and   $\Big  \langle \left (\sum_{i=1}^{N_{\mathrm{tot} }} \hat{P}_{\uparrow, i}(2g(\vec{r}_i))^2)\right)^2 \Big \rangle - \Big \langle  \sum_{i=1}^{N_{\mathrm{tot}   }} \hat{P}_{\uparrow, i}(2g(\vec{r}_i))^2  \Big \rangle^2= N_\uparrow (2 g)^4$/2.  Here, $g(\vec{r}_i)$ is the probe coupling constant for spin $i$ at position $\vec{r}_i$, and $\hat{P}_{\uparrow, i} = \ket{\uparrow_i}\bra{\uparrow_i}$ is the spin up projection operator for spin $i$.  The expectation values are evaluated for the previously stated coherent spin state, and also include averaging over the thermal distribution of atomic positions.

For our system, $2g\left(\vec{r}\right) = 2g_0 \frac{w_0}{w(z)} e^{-\frac{x^2 + y^2}{w(z)^2} } \sin(k z)$ where $w(z) = w_0 \sqrt{1 + (z/z_R)^2}$, and $w_0 = 70.6~$um is the cavity mode waist and $z_R = 1.97~$cm is the Rayleigh length calculated from the accurately measured cavity free spectral range  of $7.828(1)~$GHz and transverse mode spacings $\mathrm{TEM}_{00} - \mathrm{TEM}_{\{01, 10\}}$ of $2257(2)~$MHz and  $\mathrm{TEM}_{00} - \mathrm{TEM}_{\{02, 11, 20\}}$ of $4515(3)~$MHz measured at $795~$nm. The calculated peak coupling at the antinode is $2g_0 = 2\pi \times 607~$kHz.  Accounting for both the axial and radial averaging of the probe-atom coupling, we arrive at an effective single particle vacuum Rabi frequency of $2g = 2\pi \times 506(8)~$kHz and an effective atom number $N = 0.664 \times N_{\mathrm{tot}}$.

\subsection{Atomic Ensemble Properties}

We measured a radial temperature of $25(12)~\mu$K for the atomic ensemble by measuring the reduction in Rabi splitting versus the ballistic expansion time after turn off of the optical lattice.  The lattice was switched off on a time scale much faster than all trap frequencies. Together with the calculated average lattice depth of $U_0/h = 7.4(5)~$MHz or $360(30)~\mu$K, we infer the rms radial extent of the ensemble in the x and y directions to be $x_{rms} = y_{rms} = 10(2)~\mu$m. Assuming the axial temperature is in thermal equilibrium with the radial temperature, the rms amplitude of atomic motion in the axial direction is $z_{rms} = 24(6)~$nm.  The axial extent of the atomic ensemble is well described by a gaussian with rms width $\sigma_z = 0.84(9)~$mm, with $68\%$ of the atoms occupying $4100$ wells or $200$ atoms per well near the center of the cavity. The center of the atomic ensemble is $<1~$mm from the center of the cavity.

The trap axial and radial frequencies are $320(10)~$kHz and $0.82(3)~$kHz respectively. We sweep the probe across the $8.5(3)~$MHz FWHM of the atom-cavity resonances in $7.1$~us, averaging over $2.3$ axial oscillations. The radial motion is essentially frozen out on this time scale.  On the $200~\mu$s time scale of the entire sequence, atom-atom collisions are negligible as well.

\subsection{Probe Quantum and Technical Noise Level}

The probe vacuum noise along with technical noise contribute an uncertainty $\Delta J_{zm}$ to the estimate  of the projection $J_z$.  To calibrate this noise, we start with all atoms in \ket{\downarrow} ($\theta_B = -\pi/2$), perform a $\pi/2$ rotation, then measure the vacuum Rabi splitting twice  and label the results $\Omega_1$ and $\Omega_2$.  Each measurement may fluctuate from one trial to the next due to total atom number fluctuations, microwave power fluctuations, and projection noise; however, these sources of noise are common to both measurements within a single trial and cancel, at least to lowest order.  The measurement noise is taken to be $\Delta J_{zm} =  \Delta\left(\Omega_2^2 - \Omega_1^2\right)/(8 g^2)$.  Accounting for the effects of Raman scattering to other magnetic sublevels does not change this result.

\subsection{Added Noise from Microwave Rotations}

A classical Bloch vector can exhibit fluctuations in the measured polar angle $\theta_B$ that arise due to classical noise introduced during a rotation.  In principle, one might mistake such classical fluctuations for the quantum fluctuations due to projection noise.  To constrain the possible added-noise in $\theta_B$ due to rotations, we perform a set of auxiliary rotation and measurement sequences (summarized in Table I), each with sensitivity to different types of errors in the rotation process.  Rotation-induced noise in $\theta_B$ is distinguished from projection noise by selecting rotations that nominally return all Bloch vectors to their original orientation before  each measurement of a vacuum Rabi splitting.  The chosen rotations constrain the added noise from microwave amplitude and phase noise, and transition frequency noise (arising from the trapping potential for instance).  These noise sources are equivalent to fluctuations in the angle of rotation, and the axis of rotation.  For the two rotation sequences used to actually observe projection noise (see Table II), we estimate that the rotations contribute at most $-14(3)$~dB of added noise in $\theta_B$ relative to the predicted projection noise level for $N = 7\times 10^5$ atoms.

If the rotations are imperfect in length, then noise from the anti-squeezed quadrature can leak into the measurement quadrature.  Estimates of the imperfections in the rotation lengths constrain this noise leakage to $< -8$~dB and $< -35$~dB for the two rotation sequences of Table II.

\begin{table*}[!h]
\begin{center}
\caption{Auxiliary rotation and measurement sequences were used to constrain the magnitude of the added noise $\Delta\theta_B$ in the polar angle of the Bloch vector caused by imperfections in the rotations.  The added noise is expressed in dB relative to the predicted projection noise level at $N = 7\times 10^5$ atoms.  The atoms are initially prepared in state \ket{\downarrow} $(\theta_B=-\pi/2)$.  The rotations are parameterized as  $R\left[\psi, \phi, \theta\right]$, where $\psi$ specifies the angle of rotation about an axis, while $\phi$ and $\theta$ specify the azimuthal and polar angles of the rotation axis.  Each measurement $N_{1,2}$ corresponds to a measurement of the number of atom in \ket{\uparrow} obtained by measuring the vacuum Rabi splitting.  The fluctuations in the difference of the measurements $N _2 - N_1$ provides a measure of the fluctuations $\Delta\theta_B$.  The rotation noise sources constrained by each sequence are given in the second column, expressed to leading order in the small fluctuating parameters $\epsilon_i$.  The amplitude fluctuations $\epsilon_i$ are re-expressed in terms of  either differential $\epsilon_-$ or common-mode $\epsilon_+$ amplitude fluctuations.  The impact of microwave phase noise and atomic transition noise (for instance, due to power fluctuations in the trapping optical lattice) can be constrained on both fast and slow time scales.}  
\begin{tabular}{ | l | c | c| }
\hline
Auxiliary 			&  $\Delta \theta_B$ 		&  Measured \\
	Sequence						& 					&$ \left<  \Delta\theta_B^2 \right> \left[  \mathrm{dB} \right]$ \\
\hline
\hline

$R\left[\pi/2, 0, 0\right]$  & fast transition  & $ -16(6)$ \\
measure $N_1$  & frequency noise, & \\
 $R\left[2\pi, \pi/2, 0\right]$  & microwave phase noise  & \\		
 measure $N_2$  & & \\
\hline

$R\left[\pi/2, 0, 0\right]$  & slow and fast transition& -14(3)\\
measure $N_1$  & frequency noise, & \\
 $R\left[\pi, \pi/2, 0\right]$  &  microwave phase noise & \\		
 $R\left[\pi, -\pi/2, 0\right]$  &  & \\
 measure $N_2$  &  & \\
\hline

$R\left[(\pi/2)(1+\epsilon_1), 0, 0\right]$  & $\pi \epsilon_- $ & $-16(3)$ \\
measure $N_1$  & $\epsilon_- = \epsilon_3-\epsilon_2$ & \\
 $R\left[\pi\left(1+\epsilon_2\right), 0, 0\right]$  &   & \\		
 $R\left[\pi\left(1+\epsilon_3\right), -\pi, 0\right]$  &  & \\
 measure $N_2$  &  & \\
\hline

$R\left[(\pi/2)(1+\epsilon_1), 0, 0\right]$  & $2 \pi \epsilon_+ $ & +4(1) \\
measure $N_1$  & $\epsilon_2, \epsilon_3 \approx   \epsilon_-$ & \\
 $R\left[ \pi\left(1+\epsilon_2\right), 0, 0\right]$  &   & \\
 $R\left[ \pi\left(1+\epsilon_3\right), 0, 0\right]$  &   & \\			
 measure $N_2$  &  & \\
\hline  

\end{tabular}
\label{AuxRot}
\end{center}
\end{table*}

\begin{table*}[h]
\begin{center}
\caption{The added noise from rotations for the two pulse sequences used to observe projection noise.  Two independent rotation sequences were used to observe projection noise--each with different sensitivities to the noise sources measured with the auxiliary rotations and measurements of Table I.  The predicted added noise from each source is expressed in dB relative to the calculated projection noise for $N = 7\times 10^5$ atoms. The phases of the microwave pulses were adjusted such that $\phi_{1, 2,3} < 0.035$~ rad.}
\begin{tabular}{ | l  | c | c | c | c| }
\hline
Measurement Sequence			&   	  \multicolumn{2}{|c|}{Amplitude Noise} & Phase Noise &  Transition Noise \\
							& 	  \multicolumn{1}{c }{$\Delta \theta_B$} &   \multicolumn{1}{c |}{$\left[  \mathrm{dB} \right]$}	&$ \left[  \mathrm{dB} \right]$	&$ \left[  \mathrm{dB} \right]$ \\
\hline
\hline

$R\left[(\pi/2)\left(1+\epsilon_2\right) ,0 , 0 \right]$&	$ \pi^2 \epsilon_+^2+ \pi \epsilon_+(\phi_3 -\phi_2)$				& $<  -30$ 	& -17(3) & -20(3)\\
$R\left[(\pi/2)\left(1+\epsilon_1\right) , \pi/2 + \phi_2 , 0 \right]$ & 																	& 			&   	&\\ 
measure $ N_1 $											&  $\epsilon_1, \epsilon_2,  \epsilon_3 \approx \epsilon_+$	&			& 	&\\ 
 $R\left[\pi\left(1+\epsilon_3\right) , \pi/2 + \phi_3, 0 \right] 	$			&												&  			& 	&\\
 measure $ N_2 $											&  												&			& 	& \\
 \hline

$R\left[(\pi/2)\left(1+\epsilon_1\right) , 0 , 0 \right]$ & $\frac{\pi}{2}\phi_2^2 \epsilon_+ + \pi \epsilon_-$  							& $ -14(3)$		& $<-28$	&$ <-34$ \\
measure $ N_1 $											&  	$\epsilon_- = \epsilon_2-\epsilon_1$				&  			&	&\\ 
 $R\left[\pi\left(1+\epsilon_2\right),\phi_2, 0\right] $					&	$\epsilon_1\approx \epsilon_+$					&			&	&\\
  measure $ N_2 $											&  												&			&	& \\
\hline  
\end{tabular}
\label{NoiseConstraints}
\end{center}
\end{table*}

\end{bibunit}


\begin{thebibliography}{23}
\expandafter\ifx\csname natexlab\endcsname\relax\def\natexlab#1{#1}\fi
\expandafter\ifx\csname bibnamefont\endcsname\relax
  \def\bibnamefont#1{#1}\fi
\expandafter\ifx\csname bibfnamefont\endcsname\relax
  \def\bibfnamefont#1{#1}\fi
\expandafter\ifx\csname citenamefont\endcsname\relax
  \def\citenamefont#1{#1}\fi
\expandafter\ifx\csname url\endcsname\relax
  \def\url#1{\texttt{#1}}\fi
\expandafter\ifx\csname urlprefix\endcsname\relax\def\urlprefix{URL }\fi
\providecommand{\bibinfo}[2]{#2}
\providecommand{\eprint}[2][]{\url{#2}}

\bibitem[{\citenamefont{Ludlow et~al.}(2008)}]{LZC08}
\bibinfo{author}{\bibfnamefont{A.~D.} \bibnamefont{Ludlow}}
  \bibnamefont{et~al.}, \bibinfo{journal}{Science}
  \textbf{\bibinfo{volume}{319}}, \bibinfo{pages}{1805} (\bibinfo{year}{2008}).

\bibitem[{\citenamefont{Gustavson et~al.}(1997)\citenamefont{Gustavson, Bouyer,
  and Kasevich}}]{GBK97}
\bibinfo{author}{\bibfnamefont{T.~L.} \bibnamefont{Gustavson}},
  \bibinfo{author}{\bibfnamefont{P.}~\bibnamefont{Bouyer}}, \bibnamefont{and}
  \bibinfo{author}{\bibfnamefont{M.~A.} \bibnamefont{Kasevich}},
  \bibinfo{journal}{Phys. Rev. Lett.} \textbf{\bibinfo{volume}{78}},
  \bibinfo{pages}{2046} (\bibinfo{year}{1997}).

\bibitem[{\citenamefont{Mohr et~al.}(2008)\citenamefont{Mohr, Taylor, and
  Newell}}]{MOT08}
\bibinfo{author}{\bibfnamefont{P.~J.} \bibnamefont{Mohr}},
  \bibinfo{author}{\bibfnamefont{B.~N.} \bibnamefont{Taylor}},
  \bibnamefont{and} \bibinfo{author}{\bibfnamefont{D.~B.}
  \bibnamefont{Newell}}, \bibinfo{journal}{Rev. Mod. Phys.}
  \textbf{\bibinfo{volume}{80}}, \bibinfo{pages}{633} (\bibinfo{year}{2008}).

\bibitem[{\citenamefont{Wicht et~al.}(2002)}]{WHS02}
\bibinfo{author}{\bibfnamefont{A.}~\bibnamefont{Wicht}} \bibnamefont{et~al.},
  \bibinfo{journal}{Physica Scripta} \textbf{\bibinfo{volume}{T102}},
  \bibinfo{pages}{82} (\bibinfo{year}{2002}).

\bibitem[{\citenamefont{Andr\'e et~al.}(2004)\citenamefont{Andr\'e, S\o{}rensen,
  and Lukin}}]{ASL04}
\bibinfo{author}{\bibfnamefont{A.}~\bibnamefont{Andr\'e}},
\bibinfo{author}{\bibfnamefont{A.~S.}~\bibnamefont{S\o{}rensen}},\bibnamefont{and}
\bibinfo{author}{\bibfnamefont{M.~D.} \bibnamefont{Lukin}},
\bibinfo{journal}{Phys. Rev. Lett.} \textbf{\bibinfo{volume}{92}},
\bibinfo{pages}{230801} (\bibinfo{year}{2004}).

\bibitem[{\citenamefont{Auzinsh et~al.}(2004)}]{ABK04}
\bibinfo{author}{\bibfnamefont{M.}~\bibnamefont{Auzinsh}} \bibnamefont{et~al.},
  \bibinfo{journal}{Phys. Rev. Lett.} \textbf{\bibinfo{volume}{93}},
  \bibinfo{pages}{173002} (\bibinfo{year}{2004}).

\bibitem[{\citenamefont{Heavner et~al.}(2005)}]{HJD05}
\bibinfo{author}{\bibfnamefont{T.~P.} \bibnamefont{Heavner}}
  \bibnamefont{et~al.}, \bibinfo{journal}{Metrologia}
  \textbf{\bibinfo{volume}{42}}, \bibinfo{pages}{411} (\bibinfo{year}{2005}).

\bibitem[{\citenamefont{Lodewyck et~al.}(2009)\citenamefont{Lodewyck,
  Westergaard, and Lemonde}}]{LWL09}
\bibinfo{author}{\bibfnamefont{J.}~\bibnamefont{Lodewyck}},
  \bibinfo{author}{\bibfnamefont{P.~G.} \bibnamefont{Westergaard}},
  \bibnamefont{and} \bibinfo{author}{\bibfnamefont{P.}~\bibnamefont{Lemonde}},
  \bibinfo{journal}{Phys. Rev. A} \textbf{\bibinfo{volume}{79}},
  \bibinfo{pages}{061401} (\bibinfo{year}{2009}).

\bibitem[{\citenamefont{Leibfried et~al.}(2004)}]{LBS04}
\bibinfo{author}{\bibfnamefont{D.}~\bibnamefont{Leibfried}}
  \bibnamefont{et~al.}, \bibinfo{journal}{Science}
  \textbf{\bibinfo{volume}{304}}, \bibinfo{pages}{1476} (\bibinfo{year}{2004}).

\bibitem[{\citenamefont{Meyer et~al.}(2001)}]{MRK01}
\bibinfo{author}{\bibfnamefont{V.}~\bibnamefont{Meyer}} \bibnamefont{et~al.},
  \bibinfo{journal}{Phys. Rev. Lett.} \textbf{\bibinfo{volume}{86}},
  \bibinfo{pages}{5870} (\bibinfo{year}{2001}).

\bibitem[{\citenamefont{Monz et~al.}(2010)}]{MSB10}
\bibinfo{author}{\bibfnamefont{T.}~\bibnamefont{Monz}} \bibnamefont{et~al.},
  \bibinfo{howpublished}{e-print arXiv:quant-ph/1009.6126}
  (\bibinfo{year}{2010}).

\bibitem[{\citenamefont{Gross et~al.}(2010)}]{GZN10}
\bibinfo{author}{\bibfnamefont{C.}~\bibnamefont{Gross}} \bibnamefont{et~al.},
  \bibinfo{journal}{Nature} \textbf{\bibinfo{volume}{464}},
  \bibinfo{pages}{1165} (\bibinfo{year}{2010}).

\bibitem[{\citenamefont{Riedel et~al.}(2010)}]{RBL10}
\bibinfo{author}{\bibfnamefont{M.~F.} \bibnamefont{Riedel}}
  \bibnamefont{et~al.}, \bibinfo{journal}{Nature}
  \textbf{\bibinfo{volume}{464}}, \bibinfo{pages}{1170} (\bibinfo{year}{2010}).
  
\bibitem[{\citenamefont{Kuzmich et~al.}(1998)\citenamefont{Kuzmich, Bigelow,
  and Mandel}}]{KBM98}
\bibinfo{author}{\bibfnamefont{A.}~\bibnamefont{Kuzmich}},
  \bibinfo{author}{\bibfnamefont{N.~P.} \bibnamefont{Bigelow}},
  \bibnamefont{and} \bibinfo{author}{\bibfnamefont{L.}~\bibnamefont{Mandel}},
  \bibinfo{journal}{Europhysics Letters} \textbf{\bibinfo{volume}{42}},
  \bibinfo{pages}{481} (\bibinfo{year}{1998}).


\bibitem[{\citenamefont{Appel et~al.}(2009)}]{AWO09}
\bibinfo{author}{\bibfnamefont{J.}~\bibnamefont{Appel}} \bibnamefont{et~al.},
  \bibinfo{journal}{Proc. Natl. Acad. Sci.} \textbf{\bibinfo{volume}{106}},
  \bibinfo{pages}{10960} (\bibinfo{year}{2009}).

  \bibitem[{\citenamefont{Schleier-Smith
  et~al.}(2010)\citenamefont{Schleier-Smith, Leroux, and
  Vuleti\ifmmode~\acute{c}\else \'{c}\fi{}}}]{SLV10}
\bibinfo{author}{\bibfnamefont{M.~H.} \bibnamefont{Schleier-Smith}},
  \bibinfo{author}{\bibfnamefont{I.~D.} \bibnamefont{Leroux}},
  \bibnamefont{and}
  \bibinfo{author}{\bibfnamefont{V.}~\bibnamefont{Vuleti\ifmmode~\acute{c}\else
  \'{c}\fi{}}}, \bibinfo{journal}{Phys. Rev. Lett.}
  \textbf{\bibinfo{volume}{104}}, \bibinfo{pages}{073604}
  (\bibinfo{year}{2010}).

 \bibitem[{\citenamefont{Wasilewski et~al.}(2009)}]{WJK10}
\bibinfo{author}{\bibfnamefont{W.}~\bibnamefont{Wasilewski}} \bibnamefont{et~al.},
  \bibinfo{journal}{Phys. Rev. Lett.} \textbf{\bibinfo{volume}{104}},
  \bibinfo{pages}{133601} (\bibinfo{year}{2010}).
 
  \bibitem[{\citenamefont{Leroux et~al.}(2010)\citenamefont{Leroux,
  Schleier-Smith, and Vuleti\ifmmode~\acute{c}\else \'{c}\fi{}}}]{LSV10}
\bibinfo{author}{\bibfnamefont{I.~D.} \bibnamefont{Leroux}},
  \bibinfo{author}{\bibfnamefont{M.~H.} \bibnamefont{Schleier-Smith}},
  \bibnamefont{and}
  \bibinfo{author}{\bibfnamefont{V.}~\bibnamefont{Vuleti\ifmmode~\acute{c}\else
  \'{c}\fi{}}}, \bibinfo{journal}{Phys. Rev. Lett.}
  \textbf{\bibinfo{volume}{104}}, \bibinfo{pages}{073602}
  (\bibinfo{year}{2010}).

\bibitem[{\citenamefont{Wineland et~al.}(1994)\citenamefont{Wineland,
  Bollinger, Itano, and Heinzen}}]{WBI94}
\bibinfo{author}{\bibfnamefont{D.~J.} \bibnamefont{Wineland}},
  \bibinfo{author}{\bibfnamefont{J.~J.} \bibnamefont{Bollinger}},
  \bibinfo{author}{\bibfnamefont{W.~M.} \bibnamefont{Itano}}, \bibnamefont{and}
  \bibinfo{author}{\bibfnamefont{D.~J.} \bibnamefont{Heinzen}},
  \bibinfo{journal}{Phys. Rev. A} \textbf{\bibinfo{volume}{50}},
  \bibinfo{pages}{67} (\bibinfo{year}{1994}).

\bibitem[{OSM()}]{OSM}
\bibinfo{note}{See EPAPS Document No. [number will be inserted by publisher]
  for experimental details.}

\bibitem[{\citenamefont{Zhu et~al.}(1990)}]{ZGM90}
\bibinfo{author}{\bibfnamefont{Y.}~\bibnamefont{Zhu}} \bibnamefont{et~al.},
  \bibinfo{journal}{Phys. Rev. Lett.} \textbf{\bibinfo{volume}{64}},
  \bibinfo{pages}{2499} (\bibinfo{year}{1990}).

\bibitem[{\citenamefont{Teper et~al.}(2008)\citenamefont{Teper, Vrijsen, Lee,
  and Kasevich}}]{TVL08}
\bibinfo{author}{\bibfnamefont{I.}~\bibnamefont{Teper}},
  \bibinfo{author}{\bibfnamefont{G.}~\bibnamefont{Vrijsen}},
  \bibinfo{author}{\bibfnamefont{J.}~\bibnamefont{Lee}}, \bibnamefont{and}
  \bibinfo{author}{\bibfnamefont{M.~A.} \bibnamefont{Kasevich}},
  \bibinfo{journal}{Phys. Rev. A} \textbf{\bibinfo{volume}{78}},
  \bibinfo{pages}{051803} (\bibinfo{year}{2008}).

\bibitem[{\citenamefont{Grangier et~al.}(1998)\citenamefont{Grangier, Levenson,
  and Poizat}}]{GLP98}
\bibinfo{author}{\bibfnamefont{P.}~\bibnamefont{Grangier}},
  \bibinfo{author}{\bibfnamefont{J.~A.} \bibnamefont{Levenson}},
  \bibnamefont{and} \bibinfo{author}{\bibfnamefont{J.-P.}
  \bibnamefont{Poizat}}, \bibinfo{journal}{Nature}
  \textbf{\bibinfo{volume}{396}}, \bibinfo{pages}{537} (\bibinfo{year}{1998}).

\end{thebibliography}

\begin{thebibliography}{3}
\expandafter\ifx\csname natexlab\endcsname\relax\def\natexlab#1{#1}\fi
\expandafter\ifx\csname bibnamefont\endcsname\relax
  \def\bibnamefont#1{#1}\fi
\expandafter\ifx\csname bibfnamefont\endcsname\relax
  \def\bibfnamefont#1{#1}\fi
\expandafter\ifx\csname citenamefont\endcsname\relax
  \def\citenamefont#1{#1}\fi
\expandafter\ifx\csname url\endcsname\relax
  \def\url#1{\texttt{#1}}\fi
\expandafter\ifx\csname urlprefix\endcsname\relax\def\urlprefix{URL }\fi
\providecommand{\bibinfo}[2]{#2}
\providecommand{\eprint}[2][]{\url{#2}}


\bibitem[{\citenamefont{Leroux et~al.}(2010)\citenamefont{Leroux,
  Schleier-Smith, and Vuleti\ifmmode~\acute{c}\else \'{c}\fi{}}}]{LSV10}
\bibinfo{author}{\bibfnamefont{I.~D.} \bibnamefont{Leroux}},
  \bibinfo{author}{\bibfnamefont{M.~H.} \bibnamefont{Schleier-Smith}},
  \bibnamefont{and}
  \bibinfo{author}{\bibfnamefont{V.}~\bibnamefont{Vuleti\ifmmode~\acute{c}\else
  \'{c}\fi{}}}, \bibinfo{journal}{Phys. Rev. Lett.}
  \textbf{\bibinfo{volume}{104}}, \bibinfo{pages}{073602}
  (\bibinfo{year}{2010}).

\bibitem[{\citenamefont{Schleier-Smith
  et~al.}(2010)\citenamefont{Schleier-Smith, Leroux, and
  Vuleti\ifmmode~\acute{c}\else \'{c}\fi{}}}]{SLV10}
\bibinfo{author}{\bibfnamefont{M.~H.} \bibnamefont{Schleier-Smith}},
  \bibinfo{author}{\bibfnamefont{I.~D.} \bibnamefont{Leroux}},
  \bibnamefont{and}
  \bibinfo{author}{\bibfnamefont{V.}~\bibnamefont{Vuleti\ifmmode~\acute{c}\else
  \'{c}\fi{}}}, \bibinfo{journal}{Phys. Rev. Lett.}
  \textbf{\bibinfo{volume}{104}}, \bibinfo{pages}{073604}
  (\bibinfo{year}{2010}).

\bibitem[{\citenamefont{Appel et~al.}(2009)}]{AWO09}
\bibinfo{author}{\bibfnamefont{J.}~\bibnamefont{Appel}} \bibnamefont{et~al.},
  \bibinfo{journal}{Proc. Natl. Acad. Sci.} \textbf{\bibinfo{volume}{106}},
  \bibinfo{pages}{10960} (\bibinfo{year}{2009}).

\end{thebibliography}
\end{document}